\documentclass[10pt,conference]{IEEEtran}
\usepackage{graphicx} 
\usepackage{inputenc}
\usepackage{amsmath}
\usepackage{physics}
\usepackage{subcaption}
\usepackage{hyperref}       
\usepackage{comment}
\usepackage{fancyhdr}
\usepackage[natbib=true, sorting=none, maxcitenames=1]{biblatex}
\usepackage[capitalise]{cleveref}
\usepackage{makecell}
\title{Identifying Bottlenecks of NISQ-friendly HHL algorithms}
\author{Marc Andreu Marfany\IEEEauthorrefmark{1}\IEEEauthorrefmark{2}\IEEEauthorrefmark{3}\textsuperscript{\textsection}, Alona Sakhnenko\IEEEauthorrefmark{1}\textsuperscript{\textsection}, Jeanette Miriam Lorenz\IEEEauthorrefmark{1}\IEEEauthorrefmark{2} \\
	\IEEEauthorblockA{\IEEEauthorrefmark{1}Fraunhofer Institute for Cognitive Systems IKS,  Munich, Germany}
	\IEEEauthorblockA{\IEEEauthorrefmark{2}Ludwig-Maximilian University, Munich, Germany}
 \IEEEauthorblockA{\IEEEauthorrefmark{3}Technische Universität München, Munich, Germany}
	\texttt{\{marc.marfany.andreu,alona.sakhnenko,jeanette.miriam.lorenz\}@iks.fraunhofer.de} \\
}
\fancypagestyle{specialfooter}{%
\fancyhf{}

\fancyfoot[R]{ \noindent\fbox{%
\parbox{\textwidth}{%
{\footnotesize © 2024 IEEE.  Personal use of this material is permitted.  Permission from IEEE must be obtained for all other uses, in any current or future media, including reprinting/republishing this material for advertising or promotional purposes, creating new collective works, for resale or redistribution to servers or lists, or reuse of any copyrighted component of this work in other works.}
}
}}
}

\begin{document}

\maketitle
\begingroup\renewcommand\thefootnote{\textsection}
\footnotetext{Equal contribution}
\endgroup
\begin{abstract}
Quantum computing promises enabling solving large problem instances, e.g. large linear equation systems with HHL algorithm, once the hardware stack matures. For the foreseeable future quantum computing will remain in the so-called NISQ era, in which the algorithms need to account for the flaws of the hardware such as noise. In this work, we perform an empirical study to test scaling properties and directly related noise resilience of the the most resources-intense component of the HHL algorithm, namely QPE and its NISQ-adaptation Iterative QPE. We explore the effectiveness of noise mitigation techniques for these algorithms and investigate whether we can keep the gate number low by enforcing sparsity constraints on the input or using circuit optimization techniques provided by \texttt{Qiskit} package. Our results indicate that currently available noise mitigation techniques, such as \texttt{Qiskit readout} and \texttt{Mthree readout} packages, are insufficient for enabling results recovery even in the small instances tested here. 
Moreover, our results indicate that the scaling of these algorithms with increase in precision seems to be the most substantial obstacle. These insights allowed us to deduce an approximate bottleneck for algorithms that consider a similar time evolution as QPE. Such observations provide evidence of weaknesses of such algorithms on NISQ devices and help us formulate meaningful future research directions.
\end{abstract}

\maketitle\thispagestyle{specialfooter} 
\section{Introduction}

Many applications require solving large systems of linear equations, from portfolio optimization in finance \cite{rebentrost2018quantum} to scattering problems in electromagnetics \cite{Clader_2013}. Even Machine Learning utilizes algorithms for solving these systems in numerous ways \cite{lloyd2013quantum}. One of the most notable examples is Bayesian inference with Gaussian Processes (GP). The algorithm allows us to create uncertainty aware models that do not suffer from vanishing gradient issues. However, its scalability issues hinder the adaptation for real-world tasks \cite{BayQC}.
Classical methods for solving linear systems of equations have a polynomial cost $\mathcal{O}(N^{\omega})$, where $N$ is the size of the system and the exponent value is lower bounded by the constant, $\omega$, where $\omega\geq 2$  \cite{Complexity_LSA}. 
For instance, the Gaussian elimination algorithm, which is one of the most well known classical algorithms, has complexity~$\mathcal{O}(N^{3})$.

Quantum computing is a new paradigm of computation that leverages quantum physics phenomena to boost our computational capabilities. Many quantum algorithms, such as Grover and Shor, have proven its capacity to solve complex problems faster. This phenomenon is often referred to as \textit{quantum speed up}. The algorithm proposed by Harrow, Hassidim and Lloyd (known as HHL) \cite{Harrow_2009} solves linear system of equations in 
$\mathcal{O}(\log(N)s^2 \kappa^2 / \epsilon)$, where $\kappa$ is a conditional number of the system, $s$ is its sparsity and $\epsilon$ is the allowed error, also known as the tolerance of the solution. Unfortunately, this algorithm assumes to have access to fault-tolerant quantum computer and a Quantum Random Access Memory (QRAM). These type of devices will not be available for the years to come.

Currently, we find ourselves in the Noisy-Intermediate-Scale Quantum (NISQ) era, which implies that quantum hardware offers a limited number of noisy qubits. As of today we have up to 433-qubits machines, e.g. provided by IBM Quantum, with over 1k-qubit machines planned in the upcoming year~\cite{IBM_roadmap}. If we wish to execute fault-tolerant algorithms (such as the HHL) on NISQ machines to test their properties early-on, we need to address certain bottlenecks that arise from different subroutines that comprise the algorithms. 

In this paper, we investigate NISQ-readiness of one of the most defining subroutines of HHL - Quantum Phase Estimation (QPE). We study its properties as well as the properties of one of its most popular NISQ adaptation, Iterative QPE (IQPE) algorithm \cite{IQPE}. The contributions of this work are as follows:
\vspace{-0.25em}
\begin{enumerate}
    \item Investigation of the effects of $s$ on the scaling and performance of QPE and IQPE algorithms;
    \item Test of the effectiveness of modern noise mitigation techniques, such as \texttt{Qiskit readout} and \texttt{Mthree readout} packages, as well as circuit optimization;
    \item Study of the effects of different noise sources on QPE and IQPE algorithms;
    \item Identification of a bottleneck for utility of these algorithms that might affect a broader family of algorithms beyond the ones studied in this paper; 
\end{enumerate}
\vspace{-0.25em}
The results of this work provide insights into requirements that algorithms for solving linear systems of equations should fulfil to be more successful on NISQ machines and help define future research direction that can bring us to closer to utilizing fault-tolerant-inspired algorithms. 

\section{Related work}\label{sec:related}
Previous works have investigated the utility of the original HHL algorithm for real-world problems and built the first proof-of-concept instances to show it. \citet{BayQC} proposed to leverage the HHL algorithm to speed up the costly matrix inversion during the inference with GPs. They show the practicality of this idea on small instances of $A^{2 \times 2}$ and $A^{4 \times 4}$ matrices. Another research direction studied ways to create more NISQ-friendly HHL algorithm for specific types of problems. \citet{baskaran2023adapting} proposed an adaptation of the HHL algorithm for solving problems in quantum chemistry. They showed the effectiveness of their method on different small molecules, e.g. $LiH$ and $BeH+$, in different geometries on a 11-qubit IonQ machine. \citet{Lee_2019} propose a hybrid version of HHL algorithm that reduces the depth of the algorithms for a specific type of linear equations without a loss of solution quality. The contribution of this work lies mostly on the theoretical side with empirical evidence covering only a small $A^{2 \times 2}$ instance. \citet{Vazquez_2022} showcased the utility of Richardson extrapolation for enhancing results quality for a specific type of input (tridiagonal Toeplitz matrix). In our work, we do not constrain the type of problems for HHL, but investigate how different scarcities $s$ of $A$ effect the algorithms.

One way to adapt HHL to NISQ devices is to adapt its most expensive subroutine, such as QPE, which is the approach that we focus on in this work. \citet{IQPE} suggests an iterative way to compute the phases by utilizing a classical register for controlling the quantum gates and thus shrinking the qubit demand of the algorithm. This method is known as an \textit{iterative QPE (IQPE)} and it is in the focal point of this paper. \citet{portfol_opt} have performed a first test of IQPE-related algorithm in the HHL setup for a financial use-case with $A^{4\times 4}$ on an ion-trap device. In our work, we perform a similar study but instead of performing a test on a single use-case we investigate the properties of the algorithm across different settings. 

Other research direction developed alternative algorithms for solving large linear systems of equations specifically tailored for NISQ devices. These alternatives offer better scaling properties and better noise resilience at an expense of a speed up that original HHL approach promises. \citet{Adiabatic_QLSV} proposed an adiabatic quantum inspired approach. They avoid QPE and instead use Hamiltonian simulation. The estimated complexity of this approach is $\mathcal{O}(\kappa^2\log(\kappa)/\epsilon)$
with the time complexity being depended on the inverse of the minimum spectral gap of the Hamiltonian. Further ideas that have been presented in \cite{Berry_2014} and \cite{Truncatedtaylorevolution} targeted specifically the Hamiltonian evolution cost through, e.g., representation with truncated Taylor evolution.
\citet{Quantumwalk_inversion} suggested a matrix inversion method by performing quantum walks. Their method allows to perform a large matrix inversion ($A^{1024\times1024}$) on a real quantum device and maintains a $\mathcal{O}(N\log N)$ speed up. This algorithm, however, imposes strong requirements on the input as the spectral radius of the matrix has to be bounded. 
\citet{omalley2022nearterm} proposed solving LSE by performing the Woodbury identity and only using basic quantum subroutines like the Hadamard test or the swap test. Those routines are particularly well-suited for current hardware. The Woodbury identity is an analytical formula that describes the inverse of a matrix as a low-rank modification of another more easily invertible matrix. Their method was able to solve LSE with 16 million entries on an IBM device. \citet{Bravo_Prieto_2023} suggested a solution for LSE on a NISQ device by implementing a Variational Quantum Linear Solver (VQLS). VQLS variationally prepares the state $\ket{x}$. In general, we would like a state such that $A\ket{x}\propto \ket{b}$. In the paper, they present four different cost functions, and their quantum circuits to estimate them. They also provide a meaningful termination condition which allows to guarantee that the cost is upper bounded by $\epsilon^2 / \kappa^2$. Using the adiabatic quantum computer they perform the VQLS up to a problem size of $A^{1024\times1024}$.

Some works have performed noise studies of different quantum algorithms. \citet{oliv2022evaluating} performed a study of Variational Quantum Eigensolver (VQE) algorithm under different noisy conditions. \citet{Fontana_2021} showed higher resilience of overparametrized Ansatz for VQE algorithm under noisy conditions. In our work, we perform a similar investigation to \cite{oliv2022evaluating} with a focal point on QPE and IQPE algorithms.

\section{Background}
\subsection{Original HHL}
The HHL~\cite{Harrow_2009} algorithm is designed to solve a system of linear equations of the form
\begin{equation}
    A\textbf{x}=\textbf{b},
\end{equation}
where $A^{N \times N}$ is a Hermitian matrix of size $N \times N$, $\textbf{x}$ and \textbf{b} are of size $N$. The algorithm can be divided in three qubits sections: register $r$, clock $c$ and ancilla $a$ qubits.
The main subroutines of the algorithm \cite{step_by_step_hhl} can be summarized as follows :

\subsubsection{Encoding $\textbf{b}$}
First, we encode $\textbf{b}$ into the register qubits $r$. The number of required qubits $n_r$ depends on the embedding strategy. The most qubit efficient method and a standard choice is Amplitude embedding, which encodes $\ket{\textbf{b}} = \sum_{i=1}^{2^{n_r}} b_i\ket{i}$ and requires $n_r = \log_2(N)$ qubits and $\hat{n}_{\text{E}}$ gates. For this, we assume that we have access to some mechanism such as a QRAM, or an uploading device, or state preparation, which does not consume gates.

\subsubsection{Estimating eigenvalues of $A$}
We can encode a unitary $U = e^{iAt}$ in $r$ and, assuming $A$ is Hermitian, eigenvalues of this unitary have a form of $e^{i \lambda t}$. To extract these eigenvalues the HHL algorithm utilizes the \textit{Quantum Phase Estimation (QPE)} subroutine. The unitary $U$ is applied to $r$ qubits while being controlled by the gate on clock qubits $c$. The phase of the eigenvalue is proportional the one of $A$ \cite{step_by_step_hhl}. The clock qubits are first prepared by applying Hadamard gates, followed by the matrix time evolution (where the $U$ is applied for different times, $U^{2^{t}}$ ). The final step involves performing the Inverse Quantum Fourier Transform (IQFT) \cite{step_by_step_hhl}, which estimates the eigenvalue phase and encodes them into the clock qubits states. 

The number of required qubits is determined by 
\begin{equation}\label{eq:qubits_QPE}
    n_{\text{QPE}} = n_{\text{r}} + n_{\text{c}} = n_{\text{r}} + \lceil p +  \log_{2}(2+\frac{1}{2\epsilon})\rceil,
\end{equation}
where $p$ is the desired precision and $\epsilon$ is the allowed error. The number of gates for this subroutine is determined by the sum of gates required for its components as
\begin{equation}\label{eq:gates_QPE}
    \hat{n}_{\text{QPE}} = \hat{n}_{\text{H}} + \hat{n}_{\text{cU}} + \hat{n}_{\text{IQFT}} = n_{\text{c}} + \hat{n}_{\text{cU}} + \frac{n_{\text{c}}^2}{2}+n_{\text{c}},
\end{equation}
where $\hat{n}_{\text{H}}$ is the number of Hadamard gates, $\hat{n}_{\text{cU}}$ the number of gates for encoding $cU$ and $\hat{n}_{IQFT}$ for IQFT. The size of the custom encoding $\hat{n}_{\text{cU}}$ depends on $n_{\text{c}}$ and the quality of the input. In theoretical works this sometimes is considered a single custom gate, however, here, we need to account for transpilation into device native gates.

\subsubsection{Eigenvalue inversion}
This is done by carrying out the eigenvalue inversion on the ancilla qubit $a$ controlled by clock qubits $c$. Usually, a single qubit is selected for ancilla qubits $n_a = 1$.
The total number of rotation gates of $\hat{n}_{R}$ for this operation only depends on the number of clock qubits $\hat{n}_{R} = n_{\text{c}}$.

\subsubsection{Uncompute operation}
The operation here is to apply the Inverse QPE on $c$ and $r$. This subroutine does not require additional qubits, but it requires the same amount of gates as in \cref{eq:gates_QPE}.

\subsubsection{Measurement}
In the final step, the measurement on the ancilla and register qubits is performed, which are loaded into a classical register.

The \textit{total number of qubits} required for the HHL algorithm thus is
\begin{equation}\label{eq:HHL_width}
    n_r + n_{\text{c}} + n_a = \log_2(N) + n + \lceil \log_{2}(2+\frac{1}{2\epsilon})\rceil + 1,
\end{equation}
while the theoretical \textit{total number of gates} is
\begin{align}\label{eq:HHL_depth}
    \hat{n}_{\text{E}} + & 2 \times \hat{n}_{\text{QPE}} + \hat{n}_{R} = \nonumber \\
    & = \hat{n}_{\text{E}} + 2\times(n_{\text{c}} + \hat{n}_{\text{cU}} + \frac{n_{\text{c}}^2}{2}+n_{\text{c}}) + n_{\text{c}} \nonumber \\
    & = \hat{n}_{\text{E}} + 2 \hat{n}_{\text{cU}} + 5 n_{\text{c}} + n_{\text{c}}^2.
\end{align} 

\subsection{NISQ-adaptations of quantum algorithms}

To enable HHL on NISQ devices for big instances we need to tackle multiple issues, such as keeping the width (Eq.~\ref{eq:HHL_width}) and the depth (Eq.~\ref{eq:HHL_depth}) of the subroutines to a minimum, as well as utilize noise mitigation technique to compensate for hardware imperfection. In the following section, we are going to discuss different ideas from literature and the available technological stack that can be used to achieve this.

\subsubsection{Optimizing QPE subroutine}\label{sec:intro_IQPE}
QPE lies in the heart of the HHL algorithm and is therefore majorly responsible for its width and depth. One of the popular algorithms that reduces the width is an \textit{Iterative QPE (IQPE)} that was adapted to HHL in the NISQ scenario by \citet{portfol_opt}. This algorithm utilizes the reset and reuse principle, mid-circuit measurements and classical control bits to shrink the qubit demand (\cref{eq:qubits_QPE}) to 
\begin{equation}\label{eq:qubits_IQPE}
    n_{\text{IQPE}} = n_{\text{r}} + 1.
\end{equation}
This algorithm allows to use a single auxiliary qubit that is reset and reused to estimate the phase of an eigenvalue in an iterative fashion to acquire arbitrary precision. Additionally, it substitutes the two-qubit gates for the phase kickback, that are more error-prone, with $\hat{n}_{\text{Kickback}}$ single-qubit gates that are classically controlled. This brings the total gate number for this algorithm to 
\begin{equation}\label{eq:gates_IQPE}
    \hat{n}_{\text{IQPE}} = \hat{n}_{\text{H}} + \hat{n}_{\text{cU}} + \hat{n}_{\text{Kickback}} = \hat{n}_i + \hat{n}_{\text{cU}} + \frac{\hat{n}_{i}^2}{2} + \frac{\hat{n}_{i}}{2},
\end{equation}
where $\hat{n}_i$ is the number of iterations of IQPE. Given that $n_{\text{c}}$ and $\hat{n}_i$ fulfill a similar role of controlling precision, it is appropriate to set them to the same value. Hence, from here one we will use $n_{\text{c}}$ for IQPE as well. This brings the total counts in \cref{eq:gates_QPE} and \cref{eq:gates_IQPE} fairly close together. 

This technique has been adapted further by \citet{hybrid_solve_linear_system}, who presented a hybrid adaptation that provides more flexibility in the trade-off between width and depth of the algorithm. Another approach for modifying QPE was presented in~\cite{HHL_controlled_ry}, who suggested a method to reduce the qubit count by introducing a modular method for arbitrary controlled rotations instead of the iterative method described above. \citet{Blunt_2023} investigated an alternative adaptation of QPE known as Stochastic QPE (SQPE) on Rigetti superconducting device. SQPE changes the output of QPE algorithm type allowing for a broader spectrum of noise mitigation techniques, however, utilizing these algorithms within the HHL framework would require further work.

\subsubsection{Optimizing measurements subroutine}
An approach from a different angle was presented by \citet{Iterat_improv_hhl}, who proposed an improved iterative method for the original HHL algorithm with a reduced number of measurements. The algorithm is composed of the original HHL algorithm and an iterative process running on the classical side. The iterative process converges to a more accurate solution.

\subsubsection{Optimizing quantum circuits}
In a typical transpilation process, a quantum circuit undergoes five different stages: translation stage, layout stage, routing stage, optimisation stage and the scheduling stage. Before the optimisation stage, our circuit is decomposed into the native basis gate set of the quantum device. In this stage, extra swap gates accumulate to map the original circuit into the device topology. These processes can increase the depth and the number of gates, however, by choosing the desired level of optimisation the depth and number of gates is reduced. In \cite{Qiskit} there are four optimization levels, which are listed in \cref{tab:optimization}. The higher the level, the more methods to optimize the circuits are applied, which, however, comes at the cost of longer compilation time.

\begin{table}
    \centering
    \begin{tabular}{||c|c|c||}
        \hline
        Level & Optimisation type & Description \\
         \hline\hline
         \textbf{0} &  None & \makecell[lc]{The circuit depth and number of gates \\ is preserved as it is;} \\
         \hline
         \textbf{1} & Light & \makecell[lc]{Simple adjacent gates collapsing and \\ redundant reset removal;} \\
         \hline
         \textbf{2} & Medium & \makecell[lc]{Commutative gate cancellation and \\ redundant reset removal;} \\
         \hline
         \textbf{3} & Heavy & \makecell[lc]{Commutative gate cancellation, \\ resynthesis of two-qubit unitary blocks \\ and redundant reset removal;} \\
         \hline
    \end{tabular}
    \caption{Description of quantum circuit optimization levels during transpilation stage according to~\cite{Qiskit}.}
    \vspace{-1.5em}
    \label{tab:optimization}
\end{table}

A way to further reduce the number of gates could be to implement circuit cutting techniques like \cite{circuit_cuts}. However, by optimising and finding the cheapest answer, we might lose the computational speed up that the QPE algorithm provides (or more concretely, the speed up claimed in the overall the HHL algorithm).

\subsubsection{Adapting to NISQ noise}

Apart from reducing the width and the depth of the circuits, it is important to diminish sensitivity of the algorithm to noise in order to ensure its success on NISQ devices. Generally, depending on the origins of the noise, it can be split into coherent (e.g. miscalibration) and incoherent (e.g. environmental influence) noise \cite{Krantz_2019}. Some of the most prominent noise sources can be summarized as follows:

\begin{itemize}
    \item \textit{Readout error:}  
    With a certain probability $p$ a bit of a binary measurement flips;
    \item \textit{Reset error:} With a certain probability $p$ a qubit is reset to $\ket{1}$ state (instead of $\ket{0}$);
    \item \textit{Depolarising noise:} With a certain probability $p$ a quantum state under the influence of the environment transforms in a mixed state. 
\end{itemize}

Different techniques for mitigating noise have been developed and tested across the literature \cite{Cai_2023}. In this work, we concentrate on methods that are suitable to work on discrete distributions that QPE produce, such as a matrix-free measurement mitigation (M3) routine. 

\section{Method} \label{sec: Method}
In this work, we concentrate on investigating the utility of an NISQ-friendly IQPE method (see \cref{sec:intro_IQPE}), that is a cornerstone to other NISQ techniques, within the context of an HHL routine. We test how QPE and IQPE differ under varying circumstances, such as noise, model size and input quality, which allows us to identify bottlenecks that prohibit these algorithms to be deployed in the near future. Understanding existing challenges for these algorithms is essential for building meaningful adaptations of them in the future. In this section, we describe the technical details and architectural decisions that are involved in an experimental design.

\subsection{Models setup}
The QPE and IQPE algorithms are implemented with the \texttt{Qiskit} and \texttt{Scipy} packages. Both of these algorithms require a customised controlled gate which applies $U=e^{iAt}$ and its time evolution $t$. The unitary $U$ is generated using \texttt{linalg.expm} module from \texttt{Scipy}. This matrix is then transformed into a gate instruction using the package \texttt{UnitaryGate} from \texttt{Qiskit} for different $t$s, and control option is inserted. The rest of QPE subroutines, such as IQFT, are imported from \texttt{Qiskit} packages. For the IQPE circuit, the previously created $cU$ were used  following the design guidelines of \cite{Qiskit}. The relevant eigenvalues are associated to the taller peaks of the histogram. 

\subsection{Simulators setup}
We utilized the \texttt{Qiskit Aer} simulator for this experiment, which allows to create noiseless and tailored noisy environments. Thanks to its flexibility, we can use noise profiles of real IBM quantum machines or create custom noise models. The quantum device that we chose to simulate was \texttt{IBM Brisbane} with 127 qubits. Its respective native basis set of gates is $\{I,R_Z,SX,X\}$ for single-qubit gates and $ECR$ for two-qubit gate (see \cref{sec:brisbane_calibration} for information on device calibration).
For custom noise profiles we simulated readout, reset and depolarising noise as a discrete noisy gate model in which a noisy gate is treated as an ideal gate that is followed by a noise. We set all noise sources to a constant value except for one, which allows us to study its effects in isolation, which is made possible with \texttt{NoiseModel} package. The probabilities for the readout error is set to $p=\{10^{-i},0.5\},i=\overline{1, 5}$, for the reset error to $p=\{3\cdot10^{-i}\},i=\overline{1, 3}$ and for the depolarisation error to 
$p=\{10^{-i},3\cdot10^{-i},6\cdot10^{i}\},i=\overline{1, 6}$. 

To perform M3 mitigation technique, we use two interfaces. The first one is the sampler interface from \texttt{Qiskit Runtime}, which can be specified by setting different levels of resilience in its options. 
In the second interface, we used the \texttt{mthree} package from \cite{mthree}.

\subsection{Problem set generation}
To generate the random Positive Semi-Definite (PSD) matrix $A$, we have chosen the package \texttt{datasets} from \texttt{sklearn}, with a slight modification. In general, if we demand a high level of sparsity in a PSD matrix while maintaining a low dimension of the matrix, the output will be the identity matrix. We have modified this condition so as to output a random diagonal matrix with semi-positive elements generated with the \texttt{random} package of \texttt{numpy}. The vector \textbf{b} was also generated with the same package.

\subsection{Performance metrics}

To evaluate the results of our experiments, we evaluate the similarity of the output distribution with a uniform distribution that corresponds to a situation when the results are lost due the depolarisation of the qubits or related errors. The further the distribution is from the uniform one, the easier it is to extract the results. From our preliminary experiments, we have found that fidelity between two probability distributions captures this notion the best. The fidelity between two discrete probability distributions is defined as follows: 

\begin{equation*}
    F=(\sum_{i=0}^{2^{n-1}}\sqrt{p_{i}q_{i}})^2 = (\sum_{i=0}^{2^{n-1}}\sqrt{p_{i} / {2^{n}}})^2,
\end{equation*}
where $p_i$ is the tested probability distribution and $q_i$ is the reference distribution, which in this case is a uniform distribution.

\section{Empirical results}
In this section, we present the results from the numerical simulation. We start with comparing the performances of QPE and iterative QPE in a noiseless environment with different input conditions in Section~\ref{sec:noiseless_sim}. We continue to analyse these algorithms in a noisy simulated environment of \texttt{IBM Brisbane} device in Section~\ref{sec:general_noisy}, and then we analyse the impact of different noise sources in isolation to determine which ones the algorithms are most sensitive in Section~\ref{sec:specific_noisy}.

\subsection{Noiseless simulation}\label{sec:noiseless_sim}

\begin{figure}[h]
    \centering
\includegraphics[width=0.5\textwidth]{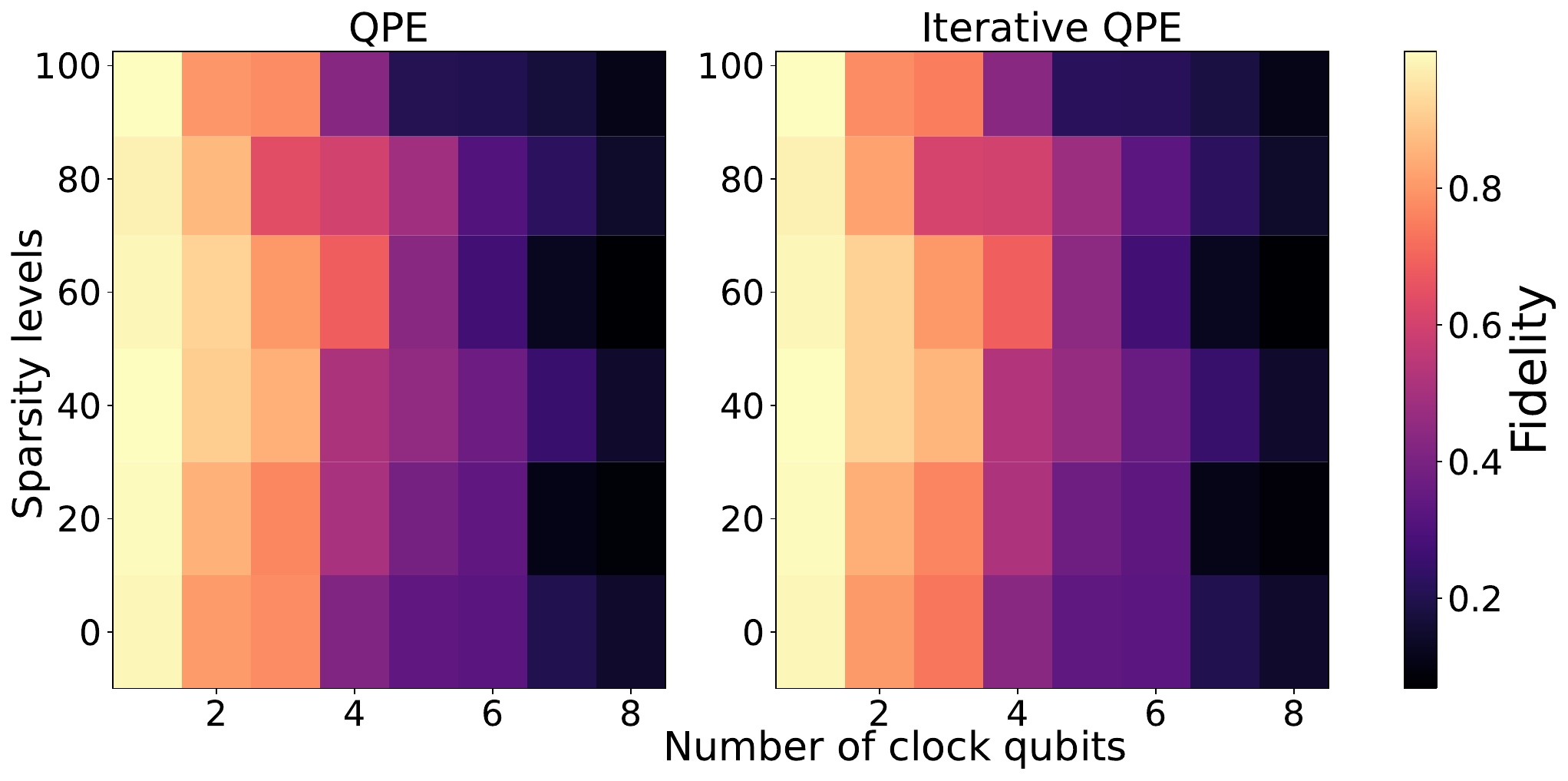}
    \caption{Fidelity heatmaps for the output distributions from the QPE and IQPE algorithms with respect to the uniform distribution in a noiseless environment. The axes correspond the number of clock qubits $n_{\text{c}}$ and the sparsity level $s$ of the input matrix $A$.}
    \label{No_noise_simulation}
\end{figure}

To begin with, the fidelities of QPE and IQPE are compared in a noiseless environment. \cref{No_noise_simulation} shows fidelity heatmaps for different numbers of clock qubits $n_{\text{c}}$ (similar to number of iterations $n_i$) and for different sparsity levels $s$ of the input matrix. For low levels of clock qubits, the fidelities are higher. 
This is due to the limited amount of bins in the resulting histograms of states. By increasing the $n_{\text{c}}$ the precision of eigenvalues will increase by $2$. Since we are increasing the resolution of these rescaled eigenvalues, the output result will continue to lower the resemblance to a flat distribution. The sparsity of the input has little effect on the fidelity in the perfect simulation environment. Only in the extreme case of almost diagonal matrix with above 80\% sparsity the fidelity slightly drops for a lower clock qubit count $n_{\text{c}}$ of around five qubits. Both algorithms perform almost identically in this environment.

\subsection{General noisy simulation} \label{sec:general_noisy}
\begin{figure}[h!]
\begin{subfigure}{0.5\textwidth}
\vspace{-1.5em}
\includegraphics[width=\textwidth]{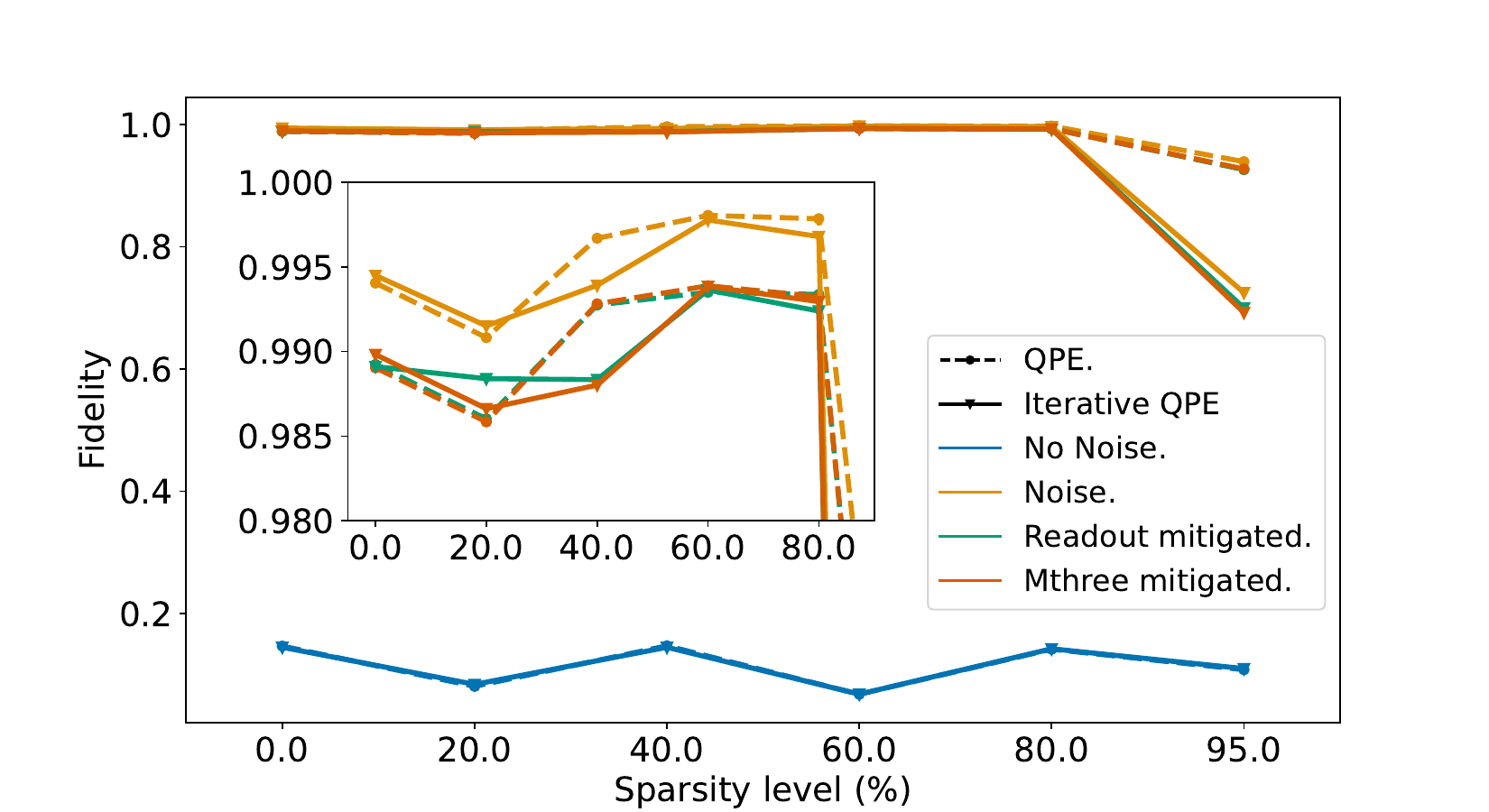}
\vspace{-1.5em}
    \caption{Comparing noise mitigation techniques, while optimization level is set to 3.}
    \label{Fidelities_optimization_3}
\end{subfigure}
\begin{subfigure}{0.5\textwidth}   \includegraphics[width=\textwidth]{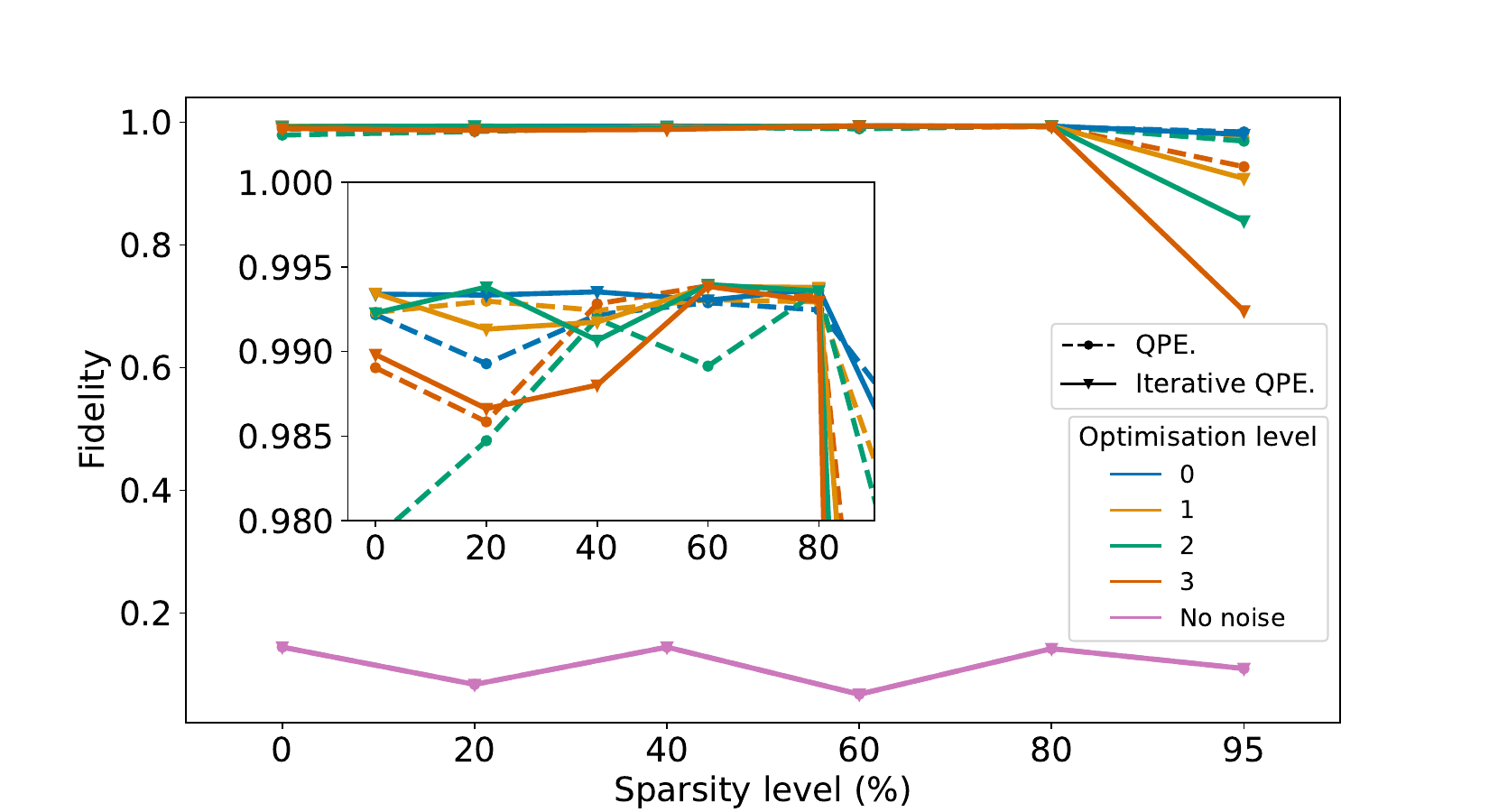}
\vspace{-1.5em}
    \caption{Comparing optimization levels, while noise mitigation is set to \texttt{mthree}.}
    \label{Fidelities_mthree}
\end{subfigure}
\caption{Fidelities of the output distribution from QPE and IQPE with respect to the sparsity level. The dimension of the matrix chosen was 8x8 and $n_{\text{c}}=8$}
\vspace{-1.5em}
\end{figure}

As the next step, we consider how QPE and IQPE compare in a noisy environment from a real quantum device \texttt{IBM Brisbane} (see \cref{sec:brisbane_calibration} for device calibration data). Given that IQPE is explicitly designed to be more suitable for NISQ devices, it is our assumption that it should be less affected by noise. So here we put to the test whether it is noise resilient enough (with the help of noise mitigation and optimization procedures) to be applied to real-world problems in the near future.

\cref{Fidelities_optimization_3} displays how the fidelities of QPE and IQPE are affected by noise under different assumptions of the input sparsity as well as the effectiveness of the different noise mitigation approaches. 
We observe that the IQPE results on average in a lower fidelity than its counter part, the QPE. This implies that it is slightly more robust against noise. Although the fidelity values are quite close, the \texttt{mthree} mitigation technique is the most efficient compared to the sampler technique. For levels of sparsity below 95\% our distribution is too flat to actually realise the true eigenvalues. A clear example can be seen in \cref{95_sparsity_b_3_nclock10}. As a matter of fact, any mitigation techniques will fail to retrieve any relevant information if the fidelity value is close to $\sim 0.997$.

\cref{Fidelities_mthree} displays how the fidelities of QPE and IQPE are affected by noise under different assumptions of the input sparsity as well as the effectiveness of the different optimization levels for the best performing mitigation technique \texttt{mthree}. We observe again that only with 95 $\%$ matrix sparsity the optimisation level of the circuit really has an impact on the fidelity. The plot shows that higher optimization level tend to show somewhat lower fidelity, with some negligible fluctuation. As mentioned above, higher optimization levels come at a higher computation cost. Given that the boost of increasing the levels was insignificant, setting the minimal level of 1 might be sufficient for most applications.

\subsection{Specific noisy simulation} \label{sec:specific_noisy}

From the previous section, we learned that QPE and IQPE algorithms struggle to perform in noisy environments, even despite deploying noise mitigation techniques. Here, we extend our study into investigating the effects of specific noise on these algorithms in isolation. Understanding the specific sensitivities of these algorithms can help us develop more meaningful mitigation techniques in the future.
\begin{figure*}[h!]
\centering
\begin{subfigure}{0.75\textwidth}
    \includegraphics[width=\textwidth]{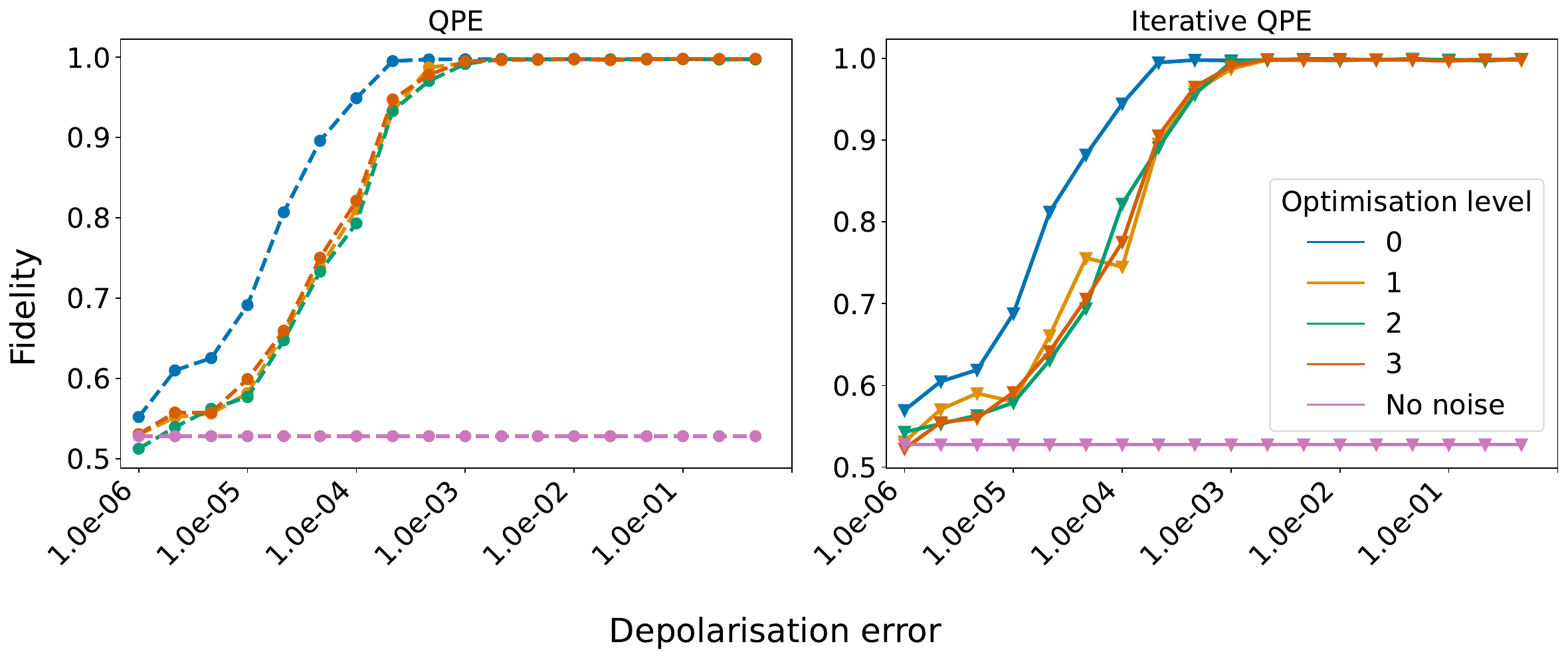}
    \caption{Readout noise probability set to $10^{-5}$ and the reset noise probability to $0.003$.}
    \label{p_r_out_0.0001_reset_0.003}
\end{subfigure}
\begin{subfigure}{0.75\textwidth}
    \includegraphics[width=\textwidth]{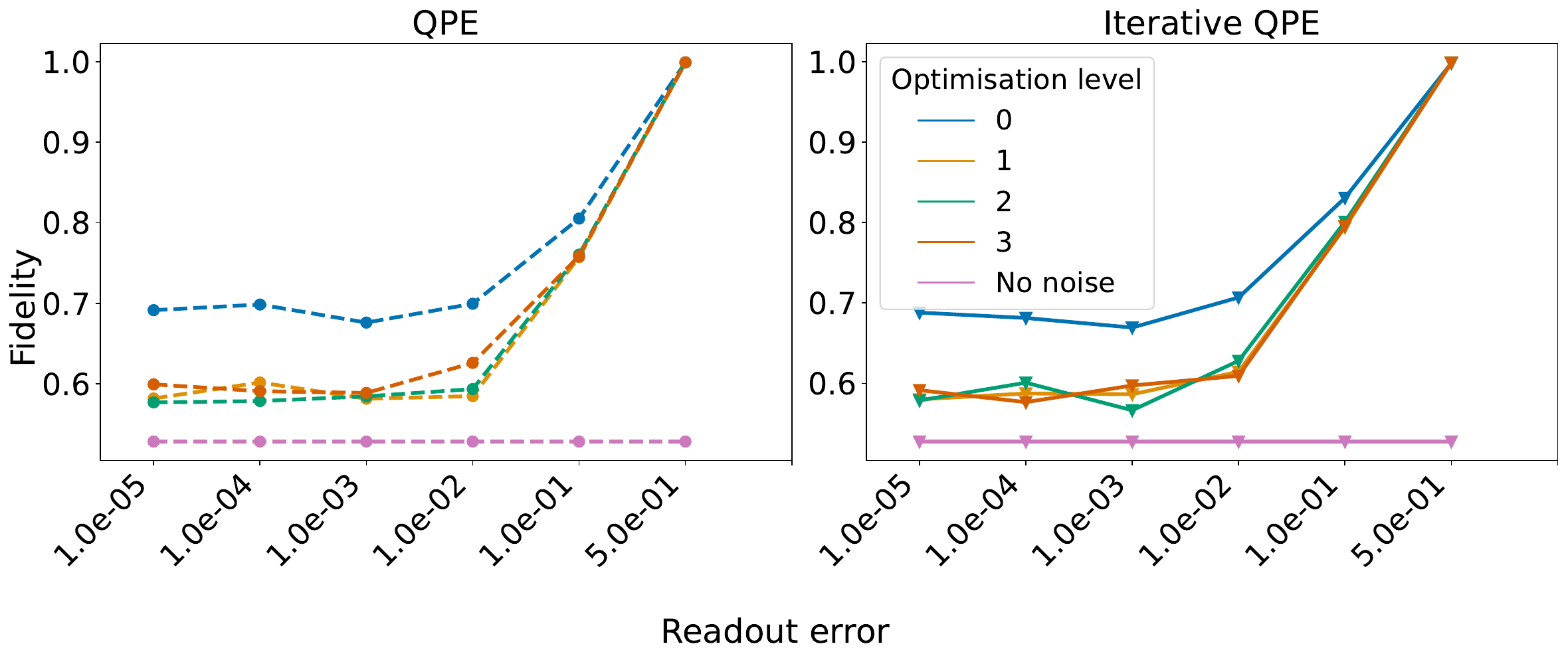}
    \caption{Depolarising noise set to $10^{-5}$ and the reset noise set to $0.003$.}
    \label{depol_noise_1e-5_p_reset_0.003}
\end{subfigure}
\begin{subfigure}{0.75\textwidth}
\includegraphics[width=\textwidth]{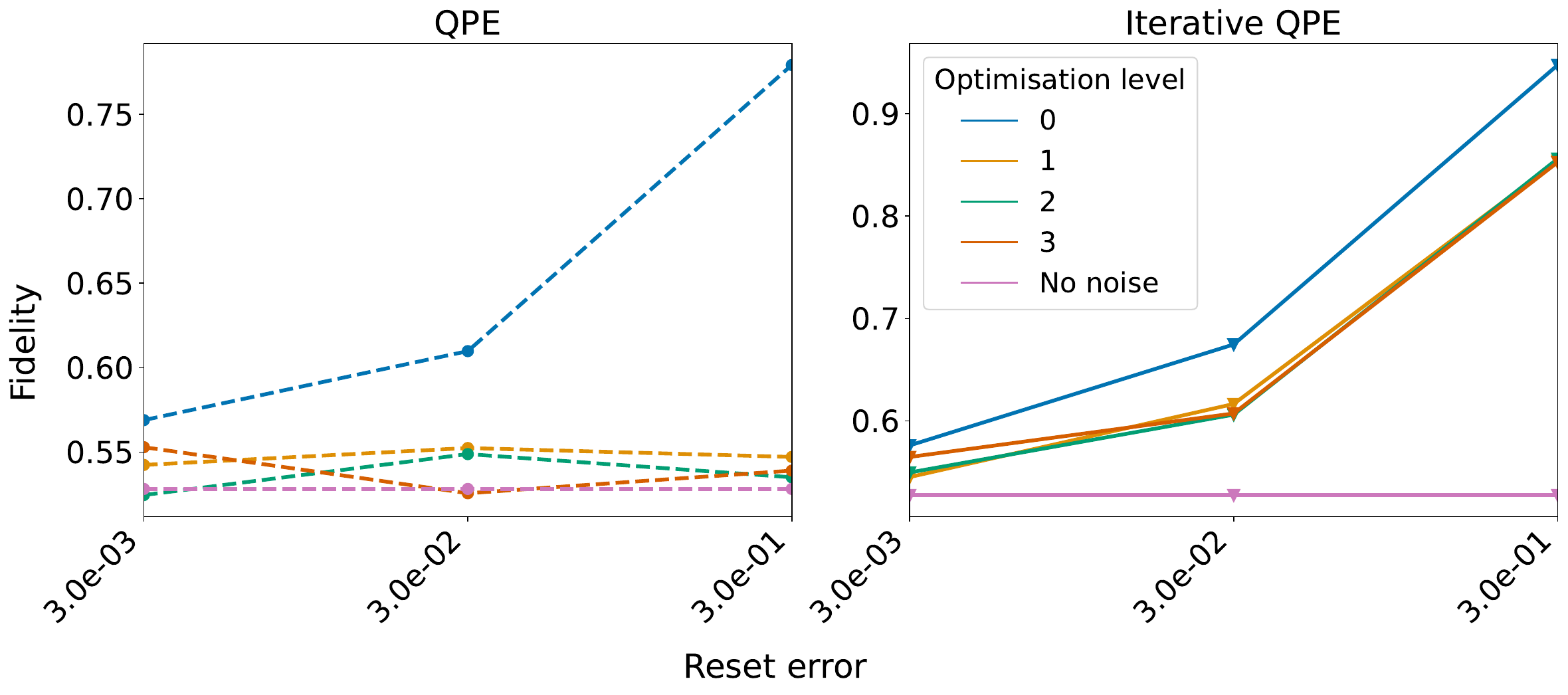}
    \caption{Depolarising noise set to $10^{-6}$ and the readout noise set to $10^{-4}$.}
    \label{depol_error_1e-6_p_r_out_1e-4}
\end{subfigure}
\caption{Fidelities of the output distribution vs the depolarising error figure \ref{p_r_out_0.0001_reset_0.003}, readout error figure \ref{depol_noise_1e-5_p_reset_0.003} and reset error figure \ref{depol_error_1e-6_p_r_out_1e-4}. In the left corresponds the QPE algorithm and on the right for the Iterative QPE. The pink line marks the fidelity for a noiseless simulation. In every case is shown the fidelity for various levels of optimization of the circuit. The dimension of the matrix was set to be 8x8 with a 35 $\%$ of sparsity level and with four clock qubits.}
\vspace{-1.5em}
\end{figure*}

\cref{p_r_out_0.0001_reset_0.003} shows how the depolarising noise affects the output. In general, we see that optimizing the circuit leads to a lower fidelity, but it is almost independent of the level. Meaning, that level 1 is already sufficient to yield most of the benefits from circuit optimization. With as little as $\sim 3\times 10^{-6}$ of noise probability, the effect on the performance of the algorithms is already palpable in the presence of noise. Circuit optimization helps to counteract the effects on noise to some extend, however, once we reach the threshold of $\sim 10^{-3}$ the results are lost. 
IQPE shows slightly lower fidelity for higher noise probabilities for optimized circuits, but the difference between optimization levels is negligible. 

From \cref{depol_noise_1e-5_p_reset_0.003} we can judge that the window of tolerance for readout noise is higher. We see that both algorithm start to rapidly increase fidelity around $\sim 10^{-2}$ noise probability. With higher noise probability IQPE loose accuracy slightly faster and at $\sim 5\times10^{-1}$ the results are lost for both algorithms.

Finally, \cref{depol_error_1e-6_p_r_out_1e-4} shows how the fidelity changes with different probabilities of the reset error. IQPE shows much higher sensitivity to reset error and, unlike QPE, the circuit optimization does not seem to counteract this effect as successfully. This behaviour of IQPE can be attributed to algorithmic structure, as IQPE has the reset instruction into its circuit while the QPE algorithm does not.

\section{Bottleneck for NISQ-friendly HHL}\label{sec:bottleneck}

\begin{figure}[h]
    \centering \includegraphics[width=0.35\textwidth]{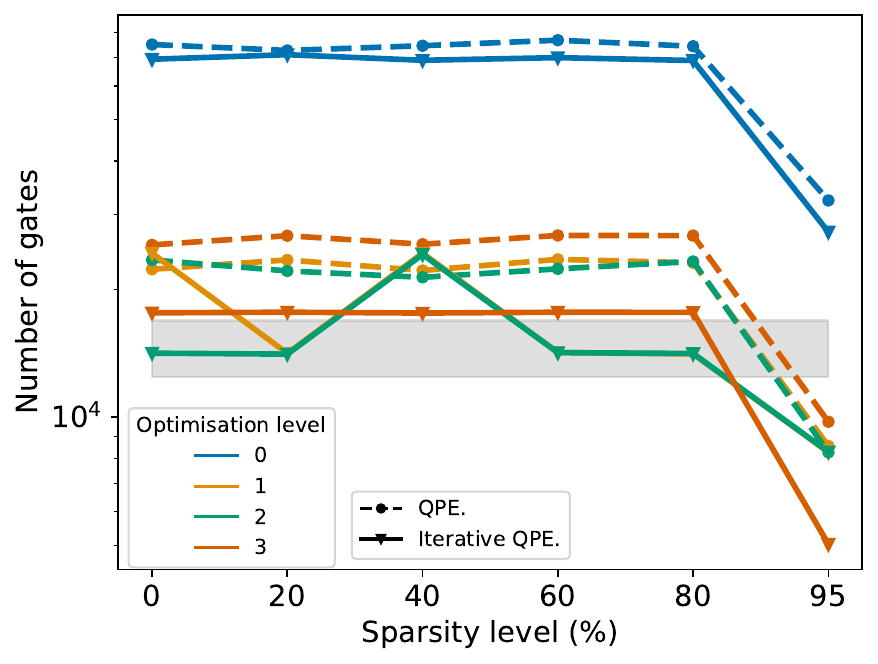}
    \caption{Total number of gates of the circuit as function of the sparsity level of the matrix A. The dimension of the matrix chosen is 8x8 and $n_{\text{c}}=8$.}
        \vspace{-1.5em}
    \label{sparsity-vs-number-gates}
\end{figure}

To understand the reasons behind a poor performance of QPE-inspired algorithms in noisy environments, it is helpful to consider the distribution of eigenvalues under different conditions. \cref{figs:histograms} shows a clear example of the loss of the relevant eigenvalues due to the noise. Notice that even though the sparsity level is selected high the results are completely lost when noise is present. If we consider that representing PSD matrices with high sparsity require approximately the same number of gates (see \cref{sparsity-vs-number-gates}),
it rejects our original hypothesis that imposing certain restrictions on the input could help when noise is involved. Our matrix input has to be near diagonal (around 95\% sparsity) for the algorithm to have a palpable drop in fidelity. This observation challenges the usefulness of this type of algorithm in the near future since a diagonal matrix is a trivial form for the inversion problem.

\begin{figure}[h!]  
\centering
\includegraphics[width=0.35\textwidth]{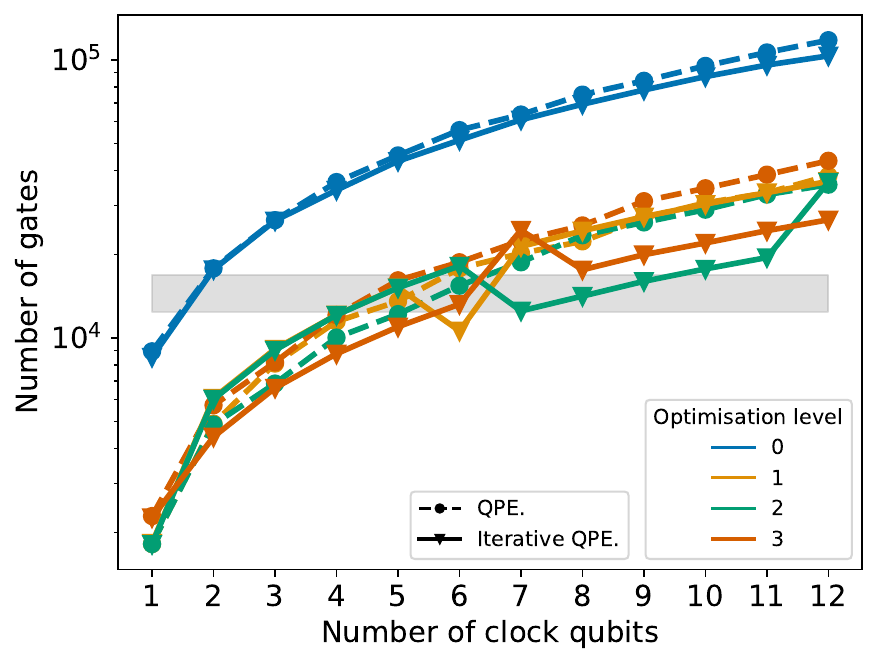}   
\caption{Number of gates (in a logarithm scale) vs  the number of times that we apply the controlled Unitary, cU. Although overlapped, it is shown the number of gates depending on the optimisation level of the circuit. The dimension of the matrix chosen was 8x8 and $n_{\text{c}}=8$. The sparsity level of the matrix is 0 $\%$.}
    \vspace{-1.5em}
\label{number_gates_sparsity_0}
\end{figure}

Shifting our attention to the number of gates, we developed a way to estimate an upper bound on the number of gates that would permit us to recover the result from the eigenvalues distributions: By closely examining the probability histogram evolution as we increase the precision ($n_{\text{c}}$ or $n_i$), we empirically estimate the upper bound of the gate number $\hat{n}_{\textbf{(I)QPE}}$ for which the error mitigation techniques remain effective. There are two approaches to find this bound for noisy environments: (1) we consider the histogram of a $A^{8\times8}$ matrix with high sparsity 95\%; (2) we consider the histogram of a smaller $A^{4\times4}$ matrix without imposing sparsity constraint. The reasoning for considering these two edge cases is that a $A^{4\times4}$ matrix demands less gates for unitary encoding and hence is less affected by noise 
However, it is also desirable to make a judgement about bigger matrices as well, such as $A^{8\times8}$, which have a tendency to resemble a uniform distribution in the presence of noise, unless the sparsity is set to 95\% (see \cref{Fidelities_optimization_3}). Separately, we need to track the number of two-qubit gates (e.g. ECR for the \texttt{IBM Brisbane}) because of the higher error rate of the two-qubit entangling gates. Specifically, the two-qubit mean error gate typically is around one order of magnitude higher than the single qubit mean error (see \cref{sec:brisbane_calibration}). When $N$ in $A^{N\times N}$ is increased, the number two-qubit entangling gates in the time evolution part rises as well. The total number of gates could stay below the threshold, but still the results are lost due to the noise. Thus, if this extra two-qubit threshold is used, we are able to formulate more reliable requirements for the algorithm.

The first estimate (1) can be obtained from  \cref{95_sparsity_b_3_nclock10}. In this case the upper bound is $\sim 4.2 \times 10^{4}$, of which $ \sim 2.8 \pm 0.5 \times10^{3}$ are two-qubit gates. 
For the second estimate (2), we need to pinpoint the threshold on $n_{\text{c}}$ after which the fidelity of the output becomes too high. For $n_{\text{c}}=8$ (see \cref{sparsity_0_b_2_nclock_8}) provides us with the empirical certification that some relevant eigenvalues are recoverable. The total number of gates is $\sim 1.4 \times 10^{4}$ with $\sim 10^{3}$ two-qubit gates. However, with $n_{\text{c}}=9$ (see \cref{sparsity_0_b_2_nclock_9}) one can not recover any relevant eigenvalue and this thus sets the upper bound at $\sim 1.7 \times 10^{4}$ and the number of two-qubit gates is $\sim 1.1 \times 10^{3}$. \footnote{This threshold is dependent on the calibration and the device.} These values highlight how low the upper bound is for the tested case.

Apart from above considerations, we need to consider the minimal number of $n_{\text{c}}$ or $n_i$ which governs the precision of the eigenvalues. Low precision correspond to less states measured, which leads to histograms with less bins meaning that each bin aggregates the states and the noise leading a rise in fidelity again (as was shown in \cref{No_noise_simulation}). The choice of precision impacts the lower bound on $n_{\text{c}}$s. Together with the upper bound on gate number discussed above, this constitutes a narrow bottleneck of utility of QPE-inspired algorithms on NISQ. One could carefully argue that NISQ-friendly algorithms (implemented on  gate based quantum computer) that consider a sort of time evolution / simulation / action like
\begin{equation}\label{eq:evolution}
    U\ket{b}=e^{i2\pi t A}\ket{b}
\end{equation}
might suffer from the same bottleneck that we encounter. This means that a vast amount of resources, will be consumed by the matrix simulation alone. 

\section{Discussion}
Our results indicate that IQPE is not yet NISQ-adjusted enough to be reliably deployed within the context of HHL for application-driven use-cases due to its scaling properties and resulting sensitivity to noise. Given the theoretical estimate from \cref{eq:gates_IQPE} that is confirmed by our empirical evidence (see \cref{number_gates_sparsity_0}), IQPE requires less gates. We have identified a narrow bottleneck of utility for this algorithm in \cref{sec:bottleneck}, which probably affects the adaptations of this algorithm (see \cref{sec:intro_IQPE}) as they
share the same core structure. 
Given the identified narrow bottleneck, the possible applicability of any algorithm with the evolution of the form as in \cref{eq:evolution} is rather challenging as this bottleneck will continue shrinking as the problem instances scale up.

This means that if one wants to scale these algorithms to larger problems (in width), a special focus has to be put onto this bottleneck. Using the \texttt{IBM Brisbane} device noise profile, we have found a threshold on the number of gates for the QPE-based algorithms (for this device). By examining the experimental data from the general noisy simulation, we deduced that the threshold in the total number of gates should be situated $\sim 1.4 \pm 0.3 \times 10^{4}$, of which $\sim 1 \pm 0.1 \times10^{3}$ constitutes the two-qubit entangling gates. The lower bound is determined by the desired precision of the eigenvalues, and together with the upper bound this constitutes the bottleneck. It is important to highlight that we considered a fairly simple use-case and a state-of-the-art device and yet the range of usefulness of an IQPE is very narrow. This bottleneck can be widened by further research into result extrapolation methods (e.g. \cite{Vazquez_2022}) and noise mitigation techniques. The results of our study of specific noise effects have indicated that QPE-inspired algorithms can tolerate the readout error the most, which is ironically the type of error targeted by noise mitigation techniques tested here. Developing noise mitigation techniques that target reset errors might be impactful for a variety of hybrid NISQ algorithm that rely on reset and reuse qubits. The noise mitigation techniques for this study were selected based on the type of algorithm output they can work with, which in this case are discrete distributions. One way to extend the family of fitting noise mitigation techniques is by performing further algorithmic changes of the QPE algorithm, e.g. \cite{Blunt_2023}, to force the algorithm to output expectation values. This adaptation, however, would require further transformation within the overall HHL algorithm. 

Our results indicate the importance of future research into two directions: (1) noise mitigation techniques that are capable of working with different types of algorithmic output or / and target specific noise sources that NISQ algorithms are particularly susceptible to and (2) algorithmic adaptations of the NISQ techniques for solving systems of linear equations that avoid time evolution. Algorithms based e.g. on quantum walks \cite{Quantumwalk_inversion} or variational approaches \cite{omalley2022nearterm} have emerged within research community, however, these algorithms do not deliver the same speed up guarantees, impose strict requirements on the input or require optimization, which might also suffer under noise \cite{Wang_2021}.

 

\section{Conclusion}
In the paper, we performed an empirical study of the properties of one of the most costly subroutines of a HHL algorithm - QPE and its more NISQ-friendly version IQPE. The properties in focus are the scaling of the algorithm depending on linear system's features and the corresponding noise resilience. Our results indicate that IQPE has a significant bottleneck defined by the upper threshold that is resulting from the unitary encoding, which controls the width and depth of the algorithm, and the lower threshold that is governed by a reasonable precision of the result. We tested this algorithm in a noisy simulation environment (with a noise profile of \texttt{IBM Brisbane}) and showed how narrow the above mentioned bottleneck is, even for a simple problem instance. We also showed that the sparsity of the system's matrix had little effect on the unitary transpilation. This work shines light on specific challenges that need to be resolved for a successful adaptation of these algorithms for NISQ devices.

\section{Acknowlegements}
mau
This research is supported by the Bavarian Ministry of Economic Affairs, Regional Development and Energy with funds from the Hightech Agenda Bayern. 

\begin{appendices}
\section{IBM Brisbane calibration data} \label{sec:brisbane_calibration}
As of the time of writing this paper, the mean errors for the \texttt{IBM Brisbane} are $8.082\times 10^{-3}$ (ECR mean error), $2.669\times 10^{-4}$ (SX mean error), $1.240\times 10^{-2}$ (Median readout error), 217.83 $\mu s$ (Median $T_1$), 129.77 $\mu s$ (Median $T_2$). This data (specifically, $T_1$ and $T_2$) has great impact on a reasonable depth of a circuit and therefore on the computed threshold.

\section{Estimating the upper bound of the gate number} \label{sec:upper_bound}
\begin{figure}[h]
\centering
\begin{subfigure}{0.45\textwidth}
\centering    \includegraphics[width=0.8\textwidth]{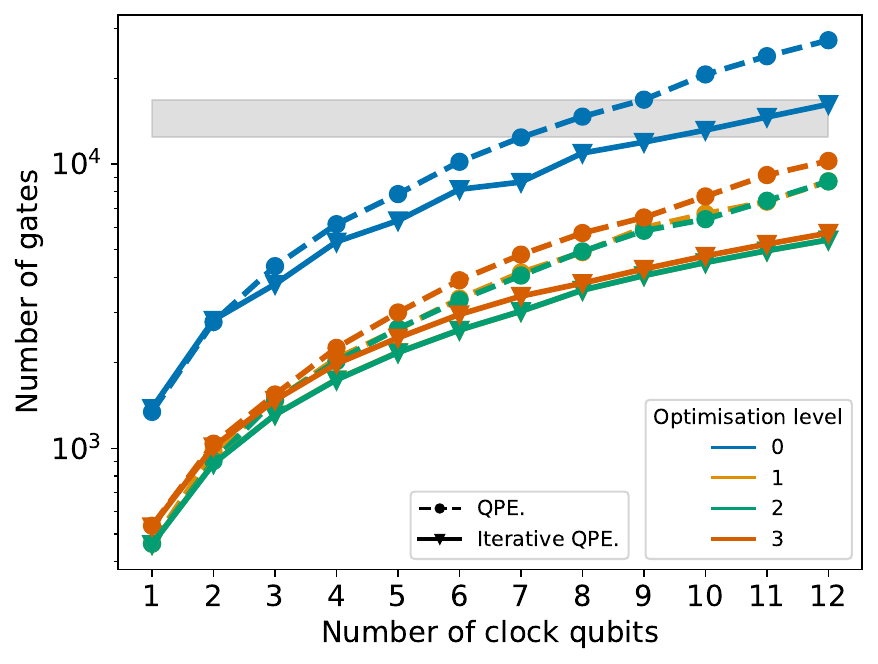}
    \caption{Single qubit gates.}
    \label{number_gates_sparsity_20}
\end{subfigure}
\begin{subfigure}{0.45\textwidth}
\centering
    \includegraphics[width=0.8\textwidth]{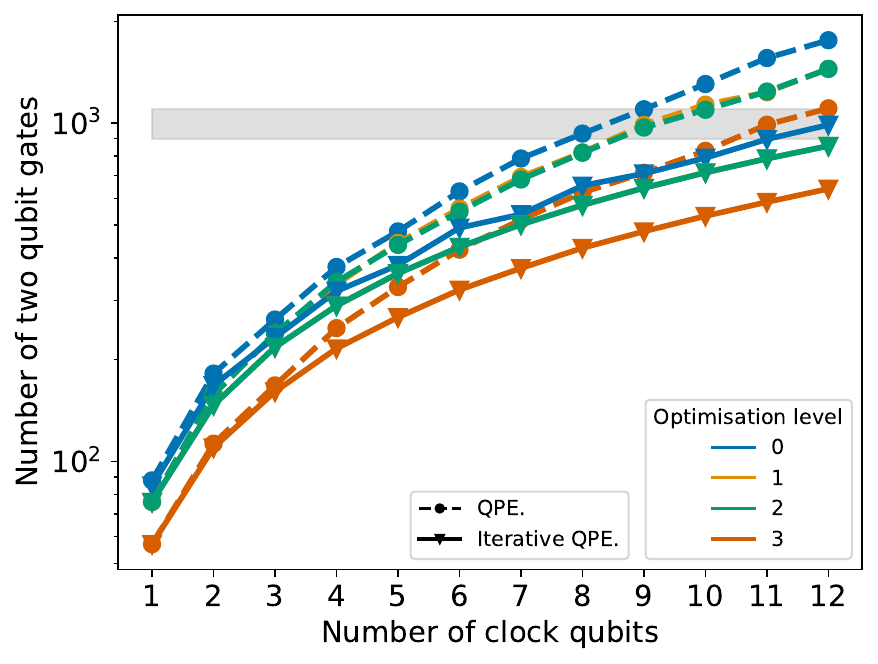}
    \caption{two-qubit gates.}
    \label{two_qubit_gates_sparsity_20}
\end{subfigure}
\caption{Number of gates (in a logarithm scale) vs $n_{\text{c}}$ for different levels of circuit optimisation. The grey area corresponds to the threshold point where above it all the results are lost.}
    \vspace{-1.5em}
\label{fig:threshold}
\end{figure}

\begin{figure}[t!]

\begin{subfigure}{0.5\textwidth}
\includegraphics[width=\textwidth]{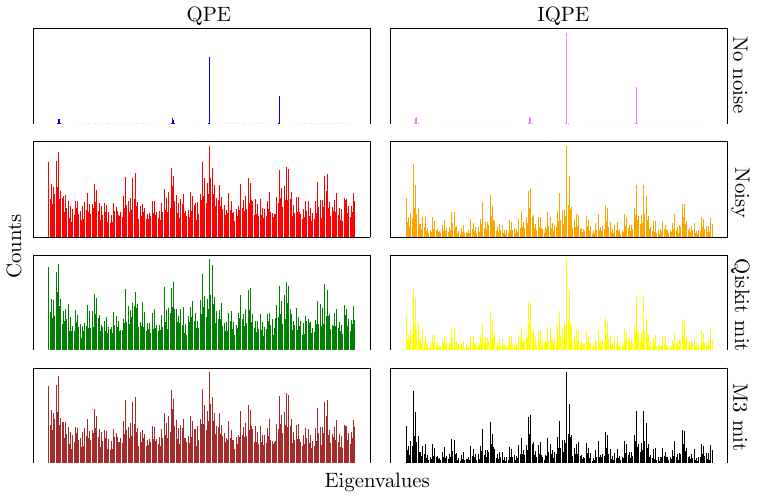}
    \caption{Histograms with recoverable eigenvalues $A^{4\times4}$ with 0 $\%$ sparsity level and $n_{\text{c}}=8$.}
    \label{sparsity_0_b_2_nclock_8}
\end{subfigure}
\begin{subfigure}{0.5\textwidth}
\includegraphics[width=\textwidth]
{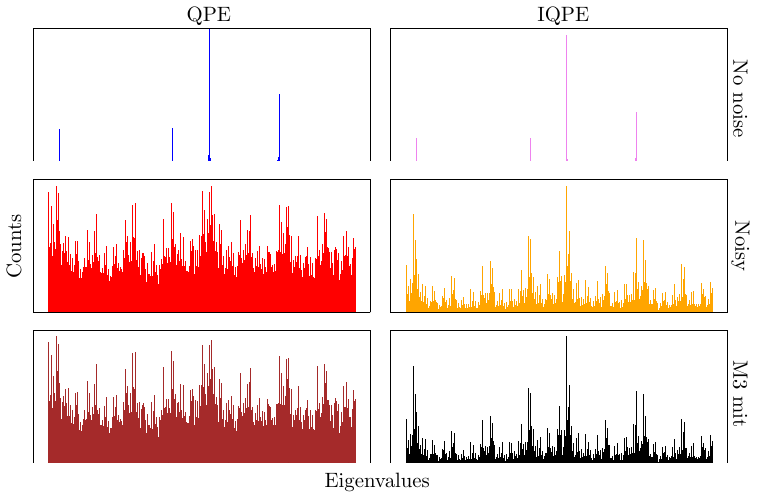}
\caption{Histograms with \textbf{non} recoverable eigenvalues $A^{4\times4}$ with 0 $\%$ sparsity level and $n_{\text{c}}=8$.}
\label{sparsity_0_b_2_nclock_9}
\end{subfigure}

\begin{subfigure}{0.5\textwidth}
    \centering
\includegraphics[width=\textwidth]{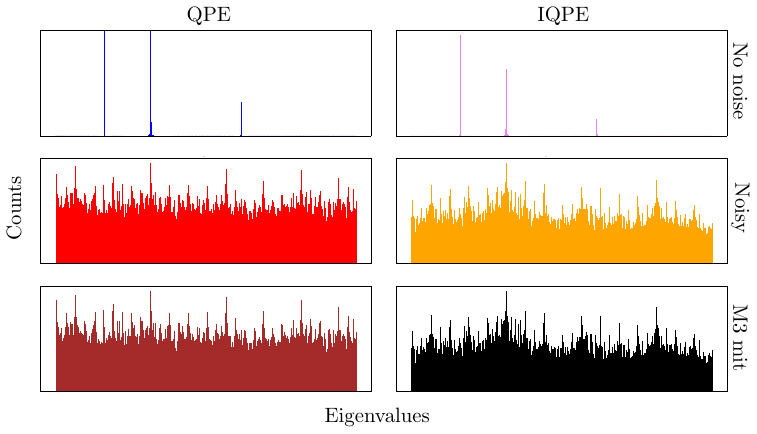}
    \caption{Histograms with \textbf{non} recoverable eigenvalues $A^{8\times8}$ with 95 $\%$ sparsity level and $n_{\text{c}}=10$.}
    \label{95_sparsity_b_3_nclock10}
\end{subfigure}
\caption{Output of QPE algorithm no level of circuit optimisation.}
\vspace{-1.5em}
\label{figs:histograms}
\end{figure}

\cref{number_gates_sparsity_20} and \cref{two_qubit_gates_sparsity_20} show the estimated upper bound for single and two-qubit gates. Single gate upper bound was estimated based of \cref{figs:histograms} as described in \cref{sec:bottleneck}. Two gate upper bound was estimated with similar consideration, but for the sake of brevity is omitted in this text.

\end{appendices}
\printbibliography
\end{document}